\documentclass[aps,pre,superscriptaddress,twocolumn]{revtex4-1}
\usepackage{bm}
\usepackage{graphicx,amsmath,latexsym,amssymb}
\usepackage{xcolor}
\usepackage{dcolumn}

\usepackage{hyperref}
\hypersetup{colorlinks=true, linkcolor=blue, citecolor=blue, urlcolor=blue}

\newcommand\blue[1]{{\color{black}#1}}

\begin{document}
\title{Topological thermalization via vortex formation in ultra-fast quenches}

\author{M. Tello-Fraile}
\affiliation{Departamento de F\'isica Te\'orica, At\'omica y \'Optica and IMUVA,  Universidad de Valladolid, Paseo Bel\'en 7, 47011 Valladolid, Spain}
\author{A. Cano}
\affiliation{Institut N\'eel, CNRS \& Univ. Grenoble Alpes, 38042 Grenoble, France}
\author{M. Donaire}
\affiliation{Departamento de F\'isica Te\'orica, At\'omica y \'Optica and IMUVA,  Universidad de Valladolid, Paseo Bel\'en 7, 47011 Valladolid, Spain}

\begin{abstract}
We investigate the thermalization of a two-component scalar field across a second-order phase transition under extremely fast quenches. 
We find that vortices start developing at the final temperature of the quench, {\it i.e.}, below the critical point. Specifically, we find that vortices emerge once the fluctuating field departures from its symmetric state and 
evolves towards a metastable and inhomogenous configuration.
The density of primordial vortices at the relaxation 
time is a decreasing function of the final temperature of the quench. 
Subsequently, vortices and antivortices annihilate at a rate that eventually determines the total thermalization time. 
This rate decreases if 
the theory contains a discrete anisotropy, which otherwise leaves the primordial vortex density unaffected.
Our results thus establish a link between the topological processes involved in the vortex dynamics and the delay in the thermalization of the system.
\end{abstract}
\maketitle


\section{Introduction}\label{Intro}

The network of topological defects left behind a phase transition is an overarching phenomenon which encodes valuable information about the underlying dynamics and the corresponding field theory in a number of systems   \cite{Lounasmaa,Dalibard,Srivastava,JoanaPedro,ArttuHindmarash,ArttuReview,ArttuKibbleDonaire,DonaireJPA,CarlosMartins,Shellard,Bullough,Zurek96,hendry94,bauerle96,donadello16,Chae12,Zurek14,Deutschlander15,PRXGriffin,PRXCano,dutta16,gerster2019dynamical}. Precisely in this context, Kibble \cite{Kibble76,Kibble80} and Zurek \cite{Zurek85} performed a seminal attempt to relate the critical dynamics 
with the spatial density of topological defects, which conforms the so-called Kibble-Zurek (KZ) mechanism. 
Their original arguments go as follows. In a finite temperature quench, as the transition proceeds, the actual correlation length of the order\blue{-}parameter distribution cannot 
forever follow its nominal equilibrium value, which diverges at the critical temperature.  
This dropout can be related to 
causality, since the speed of signal propagation remains finite. As a result, the correlation length gets frozen at some value which determines the size of the domains on which the order parameter takes approximately uniform values on the ground state manifold (i.e., 
vacuum manifold
). 
The random choice of different 
values 
at different locations of the sample is at the origin of spontaneous symmetry breaking. According to Kibble's argument \cite{Kibble76}, the evolution of the order parameter at the interface between two domains is governed by the so-called geodesic rule. That is, the order parameter interpolates through each interface following the shortest path on the vacuum manifold. In turn, this determines whether topological defects form. From this reasoning, it follows that the 
value of the correlation length 
at the freezing time determines the typical distance between topological defects.  The KZ rationale has been applied to interpret the formation of topological defects in diverse condensed-matter setups,
which has revealed intriguing departures from the postulated KZ scaling in the case of fast quenches \cite{PRXGriffin,dutta16,PRXCano,gerster2019dynamical}.

Here, we investigate numerically the formation of vortices in a two-component classical scalar field under extremely-fast-quenching conditions,  which represents a limit case of the usual KZ scenario.
This implies that 
the vortices emerge in a process that is
entirely driven by 
the dynamics of the system 
after the quench. 
In particular, the order-parameter correlations evolve according to its diffusive fluctuating dynamics at the final stationary temperature, rather than being imprinted by its critical behavior just before the freezing.  
Diffusion is caused by thermal fluctuations which, in contrast with previous 
treatments, we simulate in a self-consistent manner throughout the quenches. Ultimately, we find that
the process of formation and annihilation of vortices results in an effective
'topological delay' in the thermalization of the system. In addition, we investigate the impact of a discrete six-fold anisotropy of the system, as this applies to some condensed-matter setups of interest.

The paper is organized as follows. In Sec.\ref{sec2}  we describe the fundamentals of the model and explain our approach. In Sec.\ref{sec3} we present the results of our study. We summarize our conclusions in Sec.\ref{sec4}.

\section{Preliminaries}\label{sec2}

\subsection{The model}

We perform numerical simulations upon 
an effective Lagrangian model of a scalar field order parameter, $\mathcal{Q}$,  in which temperature fluctuations are incorporated through a Langevin term in the equation of motion. Specifically, we consider a two-component scalar order parameter, $\mathcal{Q} = (Q_1, Q_2) = (Q \cos \phi, Q \sin \phi)$, with equations of motion 
\begin{align}\label{LLLeq}
{\partial \mathcal{L} \over \partial Q_i} - \nabla {\partial\mathcal{L}   \over  \partial \nabla Q_i} = -{\partial \mathcal R \over \partial  \dot Q_i} + f_i .
\end{align}
Here $\mathcal{L} $ is the effective Lagrangian, $\mathcal R$ is the dissipative function, and  $f_i$ are the components of a stochastic Langevin force such that $\langle f_{i}(\mathbf r,t)f_{j}(\mathbf r',t')\rangle = 2 \gamma  T\delta_{ij}\delta(\mathbf r - \mathbf r ')\delta(t-t')$, with $T$ being the temperature and $\gamma$ the damping coefficient \cite{LandauLifshitz}.
In this way, we effectively consider Gaussian fluctuations of $\mathcal Q$ satisfying the fluctuation-dissipation theorem at temperature $T$ \cite{PRB_Berger}. For the sake of consistency, the effective Lagrangian is evaluated at the same temperature. This is in contrast with previous approaches ---cf. Refs.\cite{Laguna1,Laguna2}-- in which the amplitude of the noise term is kept constant and  irrespective of the value of the temperature which enters the effective Lagrangian. Further, the noise amplitude is chosen there small enough so as to hardly affect the dynamics of the phase transition. We will see later that our requirement of consistency in the Langevin term is crucial in the evolution of the transition.

\begin{figure}[b!]
   \includegraphics[width=0.45\textwidth]{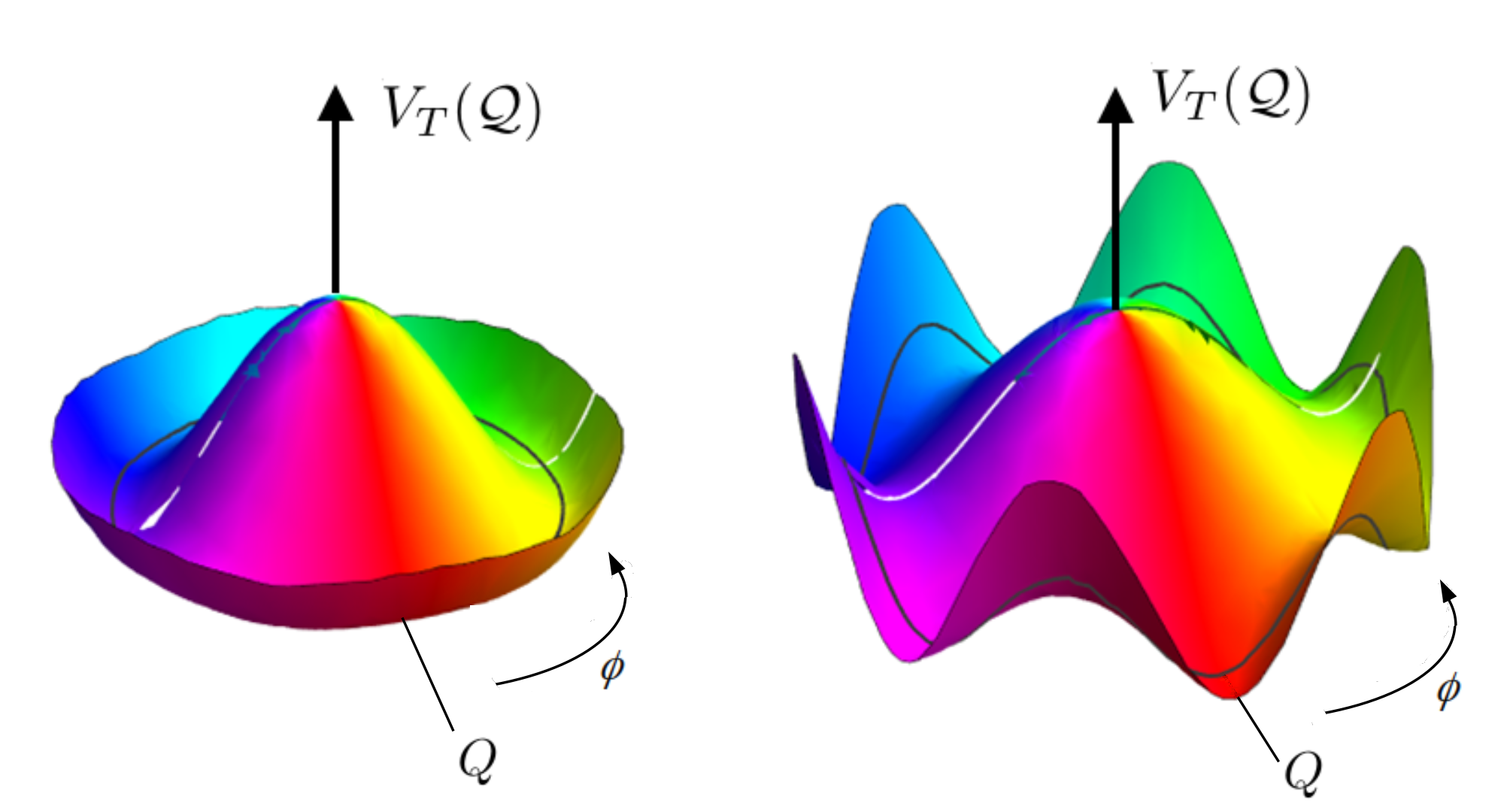}
   \caption{Sketch of the total effective potential of the order parameter in configuration space, $V_T (\mathcal{Q}) = V(Q) + V_6(\mathcal{Q})$, below the critical temperature. The field values at the minimum of $V_{T}(\mathcal{Q})$ conform the vacuum manifold. Left, potential in the weak-anisotropy regime; right, potential in the strong-anisotropy regime.}
   \label{figure_of_the_Z6_potential}
\end{figure}

In the following, we consider
an effective Lagrangian of the form
\begin{align}
\mathcal{L}  &= 
V_{T}(\mathcal{Q}) +
{g \over 2} [(\nabla Q)^2 + Q^2(\nabla \phi)^2],
\label{model}
\end{align}
where the total potential $V_{T}$ contains a $U(1)$-symmetric term, $V(Q)$, and a  six-fold anisotropy term, $V_6(\mathcal{Q})$, 
$V_{T}(\mathcal{Q}) = V(Q) + V_6(\mathcal{Q})$, 
\begin{gather}
V(Q)={a\over 2}Q^2 + {b\over 4} Q^4 + {c\over 6} Q^6, \\
V_6(\mathcal{Q})={c'\over 6} [(Q_1^2 - 3Q_2^2)Q_1]^2= {c'\over 6} Q^6\cos^2(3\phi).
\end{gather}
As is customary, we define $a = 
a_0 \epsilon$ as the control parameter with $\epsilon = (T - T_c)/T_c$ being the reduced temperature ---its temperature dependence results from the effective coupling of the system to the thermal bath.
 The coefficients $b,c,g >0$ are all constant. 
The total effective potential according to these definitions is sketched in Fig.\ref{figure_of_the_Z6_potential}.
 The critical behavior of this model belongs to the $XY$ universality class. However, the six-fold anisotropy term, $V_{6}(\mathcal{Q})$, can be tuned to describe either $U(1)$-symmetric systems like superfluids or $\mathbb{Z}_{6}$-symmetric ones like hexagonal multiferroic manganites  \cite{NatureMat14,cano-prb14,NanoLett17}. In the latter case, although the symmetry that is initially broken below $T_{c}$ is that of the $U(1)$ group, the subsequent evolution as well as the final structure of the vortices is generally affected by the $\mathbb{Z}_{6}$ anisotropy term for values of $c'$ sufficiently large. 

In static equilibrium, the uniform order parameter for which the free energy presents a minimum value must satisfy
\begin{gather}
\big[a + b Q^2 + \big(c + c' \cos^2(3\phi)\big)Q^4 \big]Q=0,
\\
Q^6 \sin (6\phi)= 0.
\end{gather}
The only solution above $T_c$ ($a>0$) is $Q_1=Q_2=0$. 
Below $T_c$ ($a<0$), the above equation presents twelve possible solutions with $Q=Q_{0}\simeq \sqrt{|a|/b}$, $\phi_n = n \pi /12$ ($n = 0, \dots, 11$). However, the actual minima of the energy correspond to either $\phi_n = n \pi /3$ if $c' <0$, or $\phi_n = (2n+1) \pi /6$ if $c' >0$, where $n = 0, \dots, 5$\blue{, unless higher-order terms are included \cite{cano-prb14}}. Thus, 
the minimum energy state of the order parameter is six-fold degenerate. Hence, the choice of any of those values by the order parameter breaks spontaneously the symmetry of the system whose phase is \blue{then} said non-symmetric.

Regarding the dynamics, we neglect inertial terms and restrict ourselves to the over-damped dynamics of $\mathcal{Q}$ introducing the dissipative function $\mathcal{R}$, \cite{LandauLifshitz}
\begin{align}
\mathcal R = {\gamma \over 2 }(\dot Q^2_1 + \dot Q^2_2) ={\gamma \over 2 }(\dot Q^2 + Q^2\dot \phi^2),
\end{align}
in accord with the Langevin force \cite{PRB_Berger}.

All in all, the time evolution of the system is described by the stochastic diffusion equation (\ref{LLLeq}) which contains a radial force $f_{Q}$
acting upon the amplitude of the order parameter, and a tangential force $f_{\phi}$ upon its phase. 
For future purposes, it is convenient to distinguish three contributions to these forces as follows,
\begin{widetext}\begin{align}
-\gamma \partial_t Q \equiv f_{Q}&
= \underbrace{a Q + b Q^3 + \big[c + c' \cos^2(3\phi)\big]Q^5}_\textrm{ radial restoring force}
\underbrace{- g\big[\nabla^2 Q - Q (\nabla \phi)^2\big]}_\textrm{radial tension force} 
\underbrace{- f_{1}\cos{\phi}-f_{2}\sin{\phi}}_\textrm{ radial stochastic force},
\label{eqc}
\\
-\gamma \partial_t \phi\equiv  f_{\phi}&=\underbrace{-c'\sin (6\phi) Q^4/6}_\textrm{tangential restoring force} 
\:\underbrace{- g\:Q^{-2}\nabla ( Q^2 \nabla \phi)}_\textrm{tangential tension force}
\:\:\underbrace{+Q^{-2}(f_{1}\sin{\phi}-f_{2}\cos{\phi})}_\textrm{tangential stochastic force}.\label{eqd}
\end{align}
\end{widetext} 
Thus, we identify a \emph{restoring force} associated to the effective potential; a \emph{tension force} associated to gradient terms; and a \emph{stochastic force} originated from the Langevin term.

\subsection{Correlation lengths and relaxation times in the mean field approximation
}\label{Meanfield}

In this section we outline the 
definition of the correlation lengths and relaxation times in the mean field approximation, as  equations can be obtained for the amplitude and phase zero modes, i.e., for uniform values of $Q$ and $\phi$, respectively.

In the symmetric phase, above $T_c$, the perturbations of the order parameter around its stable point $(0,0)$ can be written such that $(Q_1, Q_2)=(0,0) + (q_1,q_2)$. Thus, the effective Lagrangian associated to the Gaussian fluctuations is
\begin{align}
\delta \mathcal{L} &= {g\over 2} \big[ \bar{\xi}^{-2} (q_1^2 + q_2^2)  +
 (\nabla q_1)^2 + (\nabla q_2)^2 \big ],
\end{align}
where $\bar{\xi} = (g/a)^{1/2}$ is the correlation length, common to the $q_{1}$ and $q_{2}$ components. Likewise,  linearization of the equations of motion in the symmetric phase around $\mathcal{Q}=0$, neglecting the Langevin force and tensions, yields 
\begin{gather}
-\partial_t q_{1,2} = (|a|/\gamma) q_{1,2},\label{Tau}
\end{gather}
from which we identify $\bar{\tau} = \gamma/|a|$ with a relaxation time common to $q_{1}$ and $q_{2}$.

Below $T_c$, in the non-symmetric phase, it is more convenient to write $\mathcal{Q}$ in terms of $Q$ and $\phi$ instead. Hence, the value of the order parameter  around any of the six minima of the effective potential can be written now as $(Q, \phi) = (Q_0 , \phi_n) +(q,\varphi)$. In this case, the effective Lagrangian 
\blue{for} the Gaussian fluctuations reads
\begin{align}\label{FGauss}
\delta \mathcal{L} 
&= {g\over 2} \big\{ \big [ 2\bar{\xi}^{-2} q^2 + (\nabla q)^2 \big] 
+
Q^2_0 \big[ \bar{\xi}^{-2}_6 \varphi^2 + (\nabla \varphi)^2 \big]\big\}.
\end{align}
It is plain that, in addition to the correlation length of the amplitude zero mode, $\bar{\xi}$, there appears a second correlation length $\bar{\xi}_6 = [g/(3|c'|Q_0^4)]^{1/2}$ associated to the fluctuations of the phase zero mode, $\varphi$. Again, linearization around $(Q_0 , \phi_n)$ as for Eq.(\ref{FGauss}) yields the following equations for the fluctuations of the amplitude and the phase modes, respectively,
\begin{align}
- \partial_t q &= \bar{\tau}^{-1} q,\\
- \partial_t \varphi &= (3 |c'|Q_{0}^4/\gamma)\varphi,\label{Tau6}
\end{align} 
where both the Langevin force and the tension forces have been neglected. Again, $\varphi$ presents a second relaxation time,  $\bar{\tau}_6 = \gamma /(3 |c'|Q_{0}^4)$.\\


\subsection{Numerical simulations 
\label{sec:numerical}}

We perform numerical simulations on a cubic sample, considering two different situations. In the first place, we consider that the system is effectively $U(1)$-symmetric, and refer to this situation as \emph{weak-anisotropy regime}. That is, we choose the parameters in the potential $V_{T}$ such that the anisotropic term becomes irrelevant, i.e., $V_{T}(\mathcal{Q})\approx V(Q)$. In particular, for the simulations in Sec.\ref{Vorfor},  we take for the side lengths, $L_{x}=L_{y}\approx400\:l$, $L_{z}\approx120\:l$, where $l$ is the lattice spacing. The dynamics of the order parameter $\mathcal{Q}$ is governed by a discretized version of the equations (\ref{eqc}) and (\ref{eqd}). The numerical values of the parameters in those equations are $a_0= 1$, $b=2$, $c=0$, $c'=2/3$, $g=1$, $T_{c}=0.0025$, which are chosen such that  $a_{0}^{-1/2}$ becomes the unit of length scale and, at temperature $T_{c}/2$, $\epsilon=-0.5$,  $\bar{\xi}(T_{c}/2)=1$, and $\bar{\tau}(T_{c}/2)=0.1$ for $\gamma=0.1a_{0}^{1/2}$. 
In terms of the correlation lengths and relaxation times of the mean field approximation, this implies also that $\bar{\xi}\ll\bar{\xi}_{6}$, $\bar{\tau}\ll\bar{\tau}_{6}$ in the weak-anisotropy regime --cf. Table \ref{Table_Tf_n0_xi0_barxi_barxi6_tau0_bartau_bartau6} below. 

As for the situation in which $V_{6}(\mathcal{Q})$ becomes relevant, that we refer to as \emph{strong-anisotropy regime}, the coefficient $c'$ is enlarged so that the phase zero mode gets massive and the relationship between the mean field correlation lengths and relaxation times turns into $\bar{\xi}>\bar{\xi}_6$, $\bar{\tau}>\bar{\tau}_6$.  In particular, for the simulations in Sec.\ref{strongy},  we take for the side lengths $L_{x}=L_{y}=1314\:l$, $L_{z}=43\:l$. 
 The numerical values of the rest of parameters are as in the weak-anisotropy regime, except for the value of $c'$ that is taken $c'=128/3$. Simulations in the strong-anisotropy regime are performed for a single final temperature, $\epsilon_{f}=-0.3$, for which $\bar{\xi}= 1.29$, 
$\bar{\xi}_{6} = 0.59$.

Since the vortex radius scales approximately with the minimum correlation length of the mean field approximation, the lattice spacing $l$ is adapted in either case to each temperature according to the formula $l=$min$(\bar{\xi},\bar{\xi}_{6})/4$. 
It is worth mentioning  that, with the numerical values chosen for the parameters, the Ginzburg region reduces to $\epsilon \in \left[-3.96 \cdot 10^{-10}, 3.96 \cdot 10^{-10}\right]$, which is negligible in all our simulations. Therefore, effects related to the strongly nonperturbative dynamics of the order parameter within this region can be fairly discarded. Lastly, the boundary conditions imposed upon $\mathcal{Q}$ are of the kind of no-flux boundary conditions, $\mathbf{n}\cdot\nabla Q_{1,2}|_{\partial\Omega}=0$, which physically means that the polarisation vector remains fixed at the boundary $\partial\Omega$, $\mathbf{n}\perp\partial\Omega$.


In all the simulations we consider ultra-fast quenches from the symmetric phase at initial temperature $T_{i}>T_{c}$, to the non-symmetric phase at the final temperature $T_{f}<T_{c}$, assuming a uniform rate $(T_{f}-T_{i}) /\tau_{cool}$ with $\tau_{cool}$ being the cooling time.
Ultra-fast quenches are defined by the inequality $\tau_{cool}(T_{c}-T_{f})/(T_{i}-T_{f})\ll\tau_{0}$, where the term on the left hand side of this inequality is the time lapse corresponding to the temperature interval $[T_{c}, T_{f}]$, and $\tau_{0}$ is the vortex formation time measured from the time the system passes through $T_{c}$. This means that the time spent by the system  
below $T_{c}$ during the quench is negligible in comparison to the vortex formation time $\tau_{0}$. Also, it implies that the dynamics is independent  of the initial temperature. Important is to note that in our simulations the stochastic forces act on the system all the way through from the start. Finally, since the dynamics of $\mathcal{Q}$ from Eqs.(\ref{eqc}) and (\ref{eqd}) is determined by the ratio $T/\gamma$, we fix the value of the diffusion coefficient at $\gamma=0.1a_{0}^{1/2}$ in all the simulations without loss of generality.

\section{Results}\label{sec3}

\subsection{Vortex formation at weak-anisotropy}\label{Vorfor}

\begin{figure}[tb!]
\centering
    \includegraphics[width=0.475\textwidth]{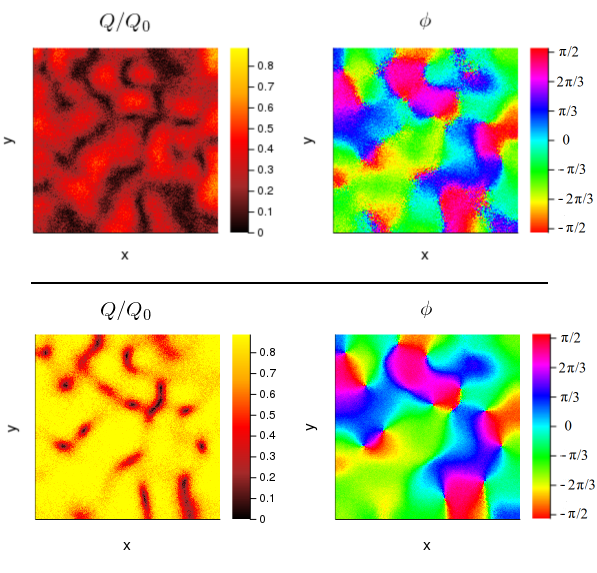}
     \caption{Snapshots of the amplitude $Q$ (left) and the phase $\phi$ (right) of the order parameter in the weak-anisotropy case for $\epsilon_f=-0.7$.  They illustrate the distribution of vortices at the formation time $\tau_{0}$ (primordial vortices, top) and at the vortex consolidation time $\tau_{1}$ (bottom).}
     \label{fig:weak-anisotropy-n}
\end{figure}

First, we analyse the effective $U(1)$-symmetric case in which the six-fold anisotropy of our model is extremely weak. That is, the case in which the phase zero mode is effectively massless and the tangential restoring force in Eq. \eqref{eqd} is negligible. The evolution of the vortex pattern obtained in this case is illustrated in Fig.~\ref{fig:weak-anisotropy-n}.

\begin{figure}[tb!]
\centering
\flushleft (a)\\   
\includegraphics[width=0.475\textwidth]{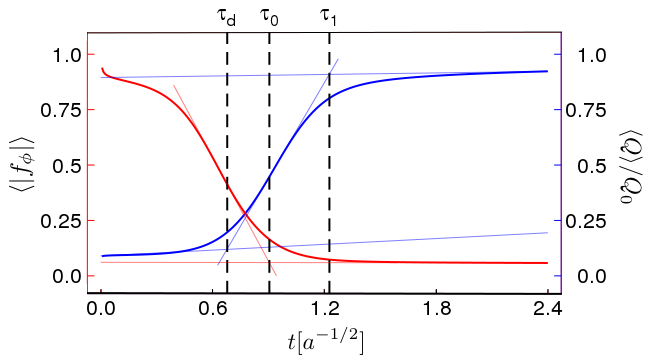}

\flushleft (b)\\
\includegraphics[width=0.45\textwidth]{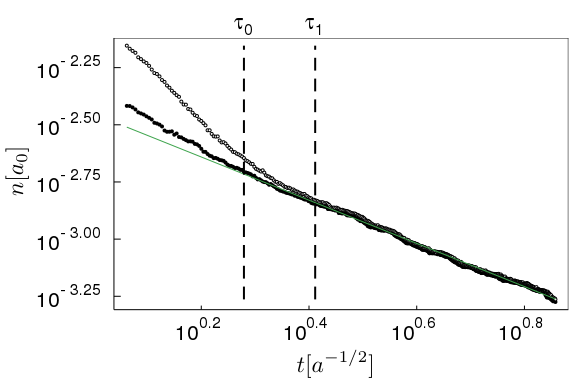}
    \caption{
    (a) Sample-averaged tangential force strength (red, in arbitrary units) and normalized amplitude of the order parameter (blue) as a function of time, in the weak-anisotropy case, for $\epsilon_f=-0.3$. The vertical dashed lines indicate the diffusion time, $\tau_{d}$, the phase relaxation time, $\tau_0$, and the vortex consolidation time, $\tau_{1}$, as defined in the main text.\\
    (b) Graphical determination of the corresponding vortex density as a function of time (log-log scale). The same power-law behavior is obtained according to two different methods (solid and open circles, respectively, cf. Ref.\cite{prepar}) beyond $\tau_1$.}
    \label{fig:weak-anisotropy}
\end{figure}

In our analysis, we first define the time $\tau_{0}$ at which primordial vortices form as the time at which a pattern of domains of well-defined uniform and stable phase $\phi$ shows up for the first time. This is the relaxation time of the phase $\phi$, and can be determined from the tangential force $f_{\phi}$ acting on $\phi$ [see Eq. \eqref{eqd}]. The behavior of the sample-averaged strength of this force $\langle|f_{\phi}|\rangle$ as a function of time is illustrated in Fig. \ref{fig:weak-anisotropy}(a) for $\epsilon_f=-0.3$. As we can see, this force drops to a small asymptotic value signaling the formation of metastable domains. 
In 
such a quasi-stationary state, only the stochastic component of Eq. (\ref{eqd}) survives. Therefore, since the anisotropic tangential restoring force term in Eq. (\ref{eqd}) is negligible, we infer that  it is the tangential tension force that causes the  relaxation of $\phi$ until domains of quasi-uniform phase get formed. This is nothing but the realization of the geodesic rule in a second order phase transition, formulated for the first time by Kibble in first order phase transitions \cite{Kibble76,DonaireJPA}. 
Thus, $\tau_{0}$ can be identified with the time at which the slope of the tangential force at its saddle point intercepts its long-time asymptote [see Fig. \ref{fig:weak-anisotropy}(a)]. 
We use this time to define the density of primordial vortices, $n_0\equiv n(\tau_{0})$, which further evolves in time as illustrated in Fig. \ref{fig:weak-anisotropy}(b). 

To further clarify the evolution of the vortex distribution we track the concomitant changes in the amplitude of order parameter. These changes reveal three dynamical regimes [see Fig.~\ref{fig:weak-anisotropy}(a)]. 
In a first stage, the linear dynamics of $Q$ is dominated by its diffusion in configuration space as a result of the stochastic force. Diffusion dominates completely the dynamics up to certain \emph{diffusion time}, $\tau_{d}$, at which $\mathcal{Q}$ starts rolling down the effective potential and non-linear effects become apparent. Diffusion causes an effective delay of the phase transition, whereas the steep descent is caused by the radial restoring force which derives from the effective potential. The phenomenon of \emph{delayed bifurcation} was firstly noticed by Lythe  in the dynamics of the phase transition of a field theory \cite{Lythe}. However, in contrast with Lythe's approach and other works \cite{Laguna1,Laguna2}, our consistent treatment of the Langevin term causes the control parameter $\epsilon$ (analogous to the bifurcation parameter in Ref. \cite{Lythe}) to take its final value much earlier than $\tau_{d}$. This condition is in fact an alternative definition for the ultra-fast character of the quench. Accordingly, the order parameter starts deviating from zero much later than the moment at which the reduced temperature has reached its final value $\epsilon_{f}$. 
Subsequently, the amplitude $Q$ eventually reaches its equilibrium value, $Q_{0}$, at any of the six minima of the effective potential, except at those points where the transition is frustrated by the presence of the topological defects. 
We identify this event with the \emph{vortex consolidation time}, $\tau_{1}$, which is slightly longer than $\tau_{0}$. $\tau_{1}$ is the relaxation time of the amplitude $Q$, and thus signals the accomplishment of the phase transition.
In fact, during the intermediate regime $Q$ is mainly driven by the radial stochastic force [see Eq.~\eqref{eqc}], while in the final stage the stochastic force takes over so that the total radial force $f_{Q}$ tends to an asymptotic stationary value. The consolidation time $\tau_{1}$ is thus associated to the crossover between these two regimes [see Fig.~\ref{fig:weak-anisotropy}(a)].

The density of primordial vortices $n_0$ as a function of the final reduced temperature $\epsilon_{f}$ is summarized in Table \ref{Table_Tf_n0_xi0_barxi_barxi6_tau0_bartau_bartau6}, together 
with the average distance between these vortices, $\xi_{0}\simeq n_{0}^{-1/2}$, relaxation times and mean field values (see Sec. \ref{Meanfield}). 
We note that the density of primordial vortices increases monotonically with the decrease of $\epsilon_{f}$ (so that the average distance between vortices decreases). The three characteristic times, in their turn, decrease as the temperature does. We interpret that these tendencies have their root in the diffusive dynamics of the order parameter in configuration space, which is enhanced by the temperature. That is, the higher the temperature, the longer the period in which the stochastic force causes $\mathcal{Q}$ to fluctuate randomly around $\mathcal{Q}=0$. Hence, $\tau_{d}$ increases with the temperature. In turn, diffusion slows down the rolling of $\mathcal{Q}$ towards any of the minima at $Q=Q_{0}$, delaying this way the start of the non-linear dynamics and thus the relaxation processes. The latter implies that $\tau_{0}$ and $\tau_{1}$ do also increase with the temperature. In addition, the persistent thermal fluctuations after the diffusion period do also enhance the periods of relaxation of the phase and the amplitude of the order parameter, causing an increase of the time intervals, $\tau_{0}-\tau_{d}$ and $\tau_{1}-\tau_{0}$,  with the temperature.  
 Altogether, it results in an effective increase of the vortex separation with the temperature. 

\begin{figure}[t!]
\centering
    \includegraphics[width=0.475\textwidth]{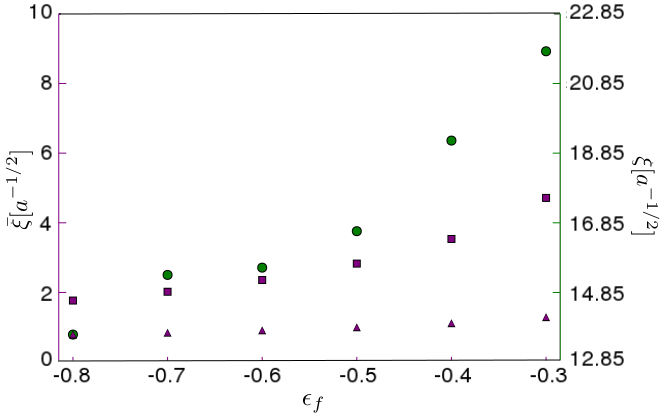}
     \caption{Average distance between vortices, $\xi_{0}$, as a function of the final temperature of the quench $\epsilon_f$ in the weak-anisotropy regime (dots), compared to the mean-field correlation lengths, $\bar{\xi}$ (triangles) and $\bar{\xi}_{6}$ (squares).} 
     
     \label{xiU1}
\end{figure}

It is also remarkable that the computed vortex correlation lengths, $\xi_{0}$, and relaxation times, $\tau_{0}$ and $\tau_{1}$, differ in an order of magnitude with respect to the values computed in the mean-field linear approximation. 
Yet, the relative variations of $\xi_{0}$, $\tau_{0}$ and $\tau_{1}$ with the temperature are approximately proportional to the variations of the mean-field correlation lengths and relaxation times, respectively, as illustrated in Fig.~\ref{xiU1} for the lengths. 
The extrapolation of this result to the standard KZ picture implies that the non-universal prefactor of the KZ scaling can play a role for the quantitative analysis of the vortex formation.
Finally, it is worth noting that, in comparison to  previous works in which the amplitude of the noise term is considered small \cite{Laguna1,Laguna2,Lythe}, the vortex density here is not affected by the relationship between that amplitude and the quench rate, but by the relationship of the final temperature and the damping rate.

\begin{table}[t!]
\begin{tabular}{ l l l l l l l }
\hline\hline
$\epsilon_f$ & $-$0.3 & $-$0.4 & $-$0.5 & $-$0.6 & $-$0.7 & $-$0.8 \\
\hline 
$n_0[a_{0}]$ & 0.0021 & 0.0027 & 0.0036 & 0.0041 & 0.0042 & 0.0054 \\
$\xi_0[a_{0}^{-1/2}]$ & 21.77 & 19.21 & 16.61 & 15.56 & 15.35 & 13.64 \\
\hline
$\bar{\xi}$ & 1.3 & 1.1 & 1.0 & 0.91 & 0.85 & 0.79 \\
$\bar{\xi}_6$ & 4.71 & 3.54 & 2.83 & 2.36 & 2.02 & 1.77 \\
\hline
$\tau_d[a_{0}^{-1/2}]$ & 1.4 & 1.1 & 0.88 & 0.77 & 0.68 & 0.62 \\
$\tau_0$ & 1.9 & 1.4 & 1.1 &  0.97 & 0.82 & 0.71 \\
$\tau_1$ & 2.61 & 1.98 & 1.59 & 1.37 & 1.19 & 1.05 \\
\hline
$\bar{\tau}$ & 0.17 & 0.13 & 0.10 & 0.083 & 0.071 & 0.063 \\
$\bar{\tau}_6$ & 2.2 & 1.3 & 0.80 & 0.56 & 0.41 & 0.31\\
\hline \hline
\end{tabular}
\caption{Compilation of the results of the numerical simulations for different final temperatures, $\epsilon_{f}$, in the weak-anisotropy regime. Lengths and times are given in units of $a_{0}^{-1/2}$, whereas the values for $n_{0}$ are in units of $a_{0}$.}
\label{Table_Tf_n0_xi0_barxi_barxi6_tau0_bartau_bartau6}
\end{table}

\subsection{Vortex formation at strong-anisotropy}\label{strongy}

Next, we investigate the impact of the six-fold anisotropy
on the formation of vortices. Thus, we consider the extended $\mathbb{Z}_6$ case described by Eq. \eqref{model} in which the phase zero mode becomes massive as described in Secs. \ref{Meanfield} and \ref{sec:numerical}. As previously mentioned, this is the situation found in hexagonal multiferroic manganites \cite{NatureMat14,cano-prb14,NanoLett17}.

Compared with the previous $U(1)$ situation, the diffusion regime is shortened [see Fig.~\ref{fig:strong-anisotropy}(a)]. This is mainly due to 
the additional contribution to radial restoring force generated by the anisotropy.
However, the phase relaxation interval extends longer due to the anisotropy contribution to the tangential restoring force that opposes to the balance of the tangential tension  (that is, $\tau_{0}-\tau_{d}$ increases).
As a result, the phase relaxation time $\tau_{0}$ is approximately equivalent in both cases [see Fig.~\ref{fig:strong-anisotropy}(a)], which further yields a similar density of primordial vortices [see Fig.~\ref{fig:strong-anisotropy}(b)]. 
Lastly, once the phase gets relaxed, it takes shorter for the vortices to consolidate under the action of the additional anisotropic radial restoring force. 
We note that, due to the increased size of the vortex cores, the sample-averaged amplitude $\langle Q\rangle$ is noticeably smaller than its equilibrium value $Q_{0}$ at all times [see Fig.~\ref{fig:strong-anisotropy}(a)]. This was previously pointed out in \cite{NanoLett17} from an experimental analysis of the static distribution of an order parameter described by the same $\mathbb{Z}_6$ model.

\subsection{Vortex network evolution}

\begin{figure}[tb!]
    \flushleft (a)\\ 
    \includegraphics[width=0.475\textwidth]{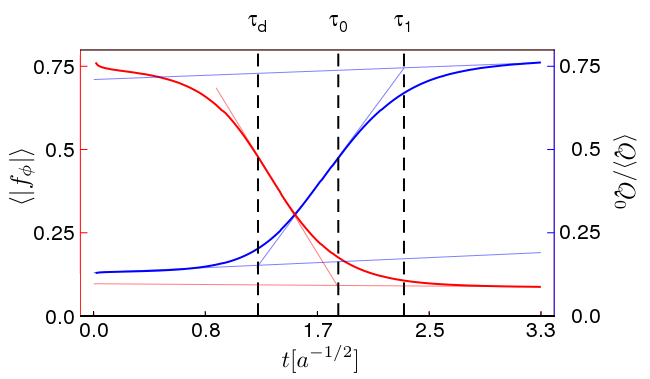}
    \flushleft (b)\\ 
     \includegraphics[width=0.45\textwidth]{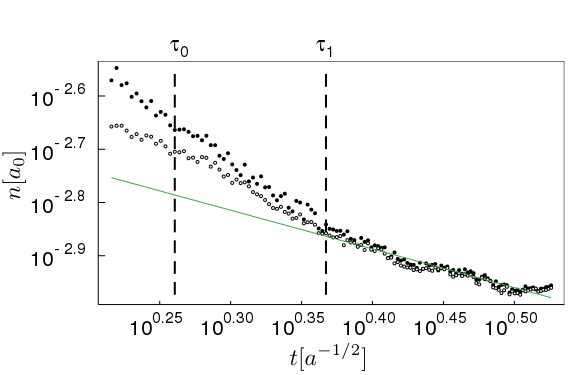}
         \caption{
    (a) Sample-averaged tangential force strength (red, in arbitrary units) and normalized amplitude of the order parameter (blue) as a function of time in the strong-anisotropy case for $\epsilon_f=-0.3$. The vertical dashed lines indicate the diffusion time, $\tau_{d}$, the phase relaxation time, $\tau_0$, and the vortex consolidation time, $\tau_{1}$, as defined in the main text.\\
    (b) Graphical determination of the corresponding vortex density as a function of time (log-log scale). The same power-law behavior is obtained according to two different methods (solid and open circles, respectively, cf. Ref.~\cite{prepar}) beyond $\tau_1$.}
     \label{fig:strong-anisotropy}
\end{figure}

Finally, we analyse the subsequent evolution of the vortex network and, in particular, the vortex-antivortex annihilation process that eventually results in a homogeneous non-symmetric phase --- see Ref.~\cite{SM} for a movie illustrating the evolution of the complete thermalization process. 
For this purpose, we fit the time evolution of the vortex density for each of the temperatures to a power-law function $n(t)\simeq n(\tau_{1})/t^{\alpha}$ for $t\geq\tau_{1}$, as shown in Figs.~\ref{fig:weak-anisotropy}(b) and \ref{fig:strong-anisotropy}(b).
Thus, we find $\alpha\approx 1$ in the $U(1)$ weak-anisotropy case and $\alpha\approx3/4$ in the $\mathbb{Z}_6$ strong-anisotropy one. 
This means that, despite the fact that the overlap between vortices is larger in the strong-anisotropy case for the vortex core is larger, the annihilation rate is slower. This signals the impact of the $\mathbb{Z}_{6}$-anisotropy in the short-range vortex-antivortex interaction.

\section{Conclusions and discussion}\label{sec4}

We have shown that, in ultra-fast quenches, the phase transition of a scalar field with a topologically non-trivial vacuum manifold, is accompanied by the formation of a metastable network of topological defects whose density  depends on the final temperature of the quench. 

The formation process involves three distinguishable and complementary mechanisms. Namely, the diffusive dynamics of the order parameter in configuration space as a result of its coupling to the thermal bath;  the local relaxation of its phase as a result of the tension forces between adjacent domains; and the global relaxation of the amplitude of the order parameter as it rolls down the effective potential. Each of these effects posses characteristic times, $\tau_{d}$, $\tau_{0}$, and $\tau_{1}$, respectively. The primordial vortex network shows up at $\tau_{0}$, whereas its consolidation takes place at $\tau_{1}$. Hence, the spatial distribution of vortices is not determined by their core radius, but by the correlation length of the phase of the order parameter at $\tau_{0}$, $\xi_{0}$. \blue{
While the alternative choice of $\tau_{1}$ as the characteristic time for this determination is also possible, it is less physical since then the intertwined dynamics of both the phase and the amplitude have an impact on the resulting topological structure. 
In fact, from a practical perspective, the unambiguous connection between the phase relaxation time and the density of primordial vortices is one of main results of our work.}

We find that the distance between the primordial vortices increases monotonically with the final temperature of the quench. We explain this behaviour as a result of the persistent thermal fluctuations of the order parameter throughout the transition.  Hence, not only $\tau_{d}$ but also the intervals between times $\tau_{d}$, $\tau_{0}$, and $\tau_{1}$ increase with the temperature, causing an effective delay of the phase transition as well as an increase on the average distance between vortices. 

In the weak-anisotropy regime, once formed, vortices and antivortices annihilate at a rate inversely proportional to time. In turn, all these effects hinder the dissipation of energy, causing a delay in the accomplishment of the thermalization process. 
Including a six-fold anisotropy we find that, whereas the vortex formation process is unaffected, their annihilation rate slows down, signaling the impact of the $\mathbb{Z}_{6}$-anisotropy in the short-range vortex-antivortex interaction.

\acknowledgments
This work was supported by the Spanish grants
MTM2014-57129-C2-1-P (MINECO), BU229P18 and VA137G18 (JCyL). We have made use of the strong performance computing resources of the Castilla y Le\'on Supercomputing Center (SCAYLE, www.scayle.es), financed by the European Regional Development Fund (ERDF). M.T.F. acknowledges financial support from the European Social Fund, the Operational Programme of Junta de Castilla y Le\'on and the regional Ministry of Education.

\end{document}